\documentclass[fleqn,usenatbib]{mnras}

\usepackage{newtxtext,newtxmath}
\usepackage{pdflscape}

\usepackage[T1]{fontenc}

\DeclareRobustCommand{\VAN}[3]{#2}
\let\VANthebibliography\thebibliography
\def\thebibliography{\DeclareRobustCommand{\VAN}[3]{##3}\VANthebibliography}

\usepackage{scalerel}
\usepackage{tikz}
\usetikzlibrary{svg.path}

\definecolor{orcidlogocol}{HTML}{A6CE39}
\tikzset{
  orcidlogo/.pic={
    \fill[orcidlogocol] svg{M256,128c0,70.7-57.3,128-128,128C57.3,256,0,198.7,0,128C0,57.3,57.3,0,128,0C198.7,0,256,57.3,256,128z};
    \fill[white] svg{M86.3,186.2H70.9V79.1h15.4v48.4V186.2z}
                 svg{M108.9,79.1h41.6c39.6,0,57,28.3,57,53.6c0,27.5-21.5,53.6-56.8,53.6h-41.8V79.1z M124.3,172.4h24.5c34.9,0,42.9-26.5,42.9-39.7c0-21.5-13.7-39.7-43.7-39.7h-23.7V172.4z}
                 svg{M88.7,56.8c0,5.5-4.5,10.1-10.1,10.1c-5.6,0-10.1-4.6-10.1-10.1c0-5.6,4.5-10.1,10.1-10.1C84.2,46.7,88.7,51.3,88.7,56.8z};
  }
}

\newcommand\orcidicon[1]{\href{https://orcid.org/#1}{\mbox{\scalerel*{
\begin{tikzpicture}[yscale=-1,transform shape]
\pic{orcidlogo};
\end{tikzpicture}
}{|}}}}
\usepackage{hyperref}
\usepackage{graphicx}	
\usepackage{amsmath}	

\newcommand\msun{M_{\odot}}
\newcommand\mhi{M_{\ion{H}{i}}}
\newcommand\LV{L_{V}}
\newcommand\lsun{L_{\odot}}

\newcommand\wfty{W_{50}}
\newcommand\wftyc{W_{50,c}}
\newcommand{\kms}{{\ensuremath{\mathrm{km\,s^{-1}}}}}
\newcommand\mhilim{M^{lim}_{\ion{H}{i}}}
\defcitealias{Carlsten2020}{C20}
\defcitealias{Carlsten2021}{C21}

\title[\ion{H}{i} in LV Satellites]{\ion{H}{i} Properties of Satellite Galaxies around Local Volume Hosts}

\author[Karunakaran et al.]{Ananthan Karunakaran$^{1\,\orcidicon{0000-0001-8855-3635}}$\,\thanks{E-mail: ananthan@iaa.es}, Kristine Spekkens$^{2,3\,\orcidicon{0000-0002-0956-7949}}$, Rhys Carroll$^{2}$, David J. Sand$^{4\,\orcidicon{0000-0003-4102-380X}}$,
\newauthor{Paul Bennet$^{5\,\orcidicon{0000-0001-8354-7279}}$, Denija Crnojevi\'{c}$^{6\,\orcidicon{0000-0002-1763-4128}}$, Michael G. Jones$^{4\,\orcidicon{0000-0002-5434-4904}}$,Bur\c{c}{\rlap{\.}\i}n Mutlu-Pakd{\rlap{\.}\i}l$^{7,8\,\orcidicon{0000-0001-9649-4815}}$}
\\
$^{1}$Instituto de Astrof\'{i}sica de Andaluc\'{i}a (CSIC), Glorieta de la Astronom\'{i}a, 18008 Granada, Spain\\
$^{2}${Department of Physics and Space Science, Royal Military College of Canada P.O. Box 17000, Station Forces Kingston, ON K7K 7B4, Canada}\\
$^{3}${Department of Physics, Engineering Physics and Astronomy, Queen's University, Kingston, ON K7L 3N6, Canada}\\
$^{4}$Steward Observatory, University of Arizona, 933 North Cherry Avenue, Rm. N204, Tucson, AZ 85721-0065, USA\\
$^{5}${Space Telescope Science Institute, 3700 San Martin Drive, Baltimore, MD 21218, USA}\\
$^{6}${University of Tampa, 401 West Kennedy Boulevard, Tampa, FL 33606, USA}\\
$^{7}${Kavli Institute for Cosmological Physics, University of Chicago, Chicago, IL 60637, USA}\\
$^{8}${Department of Astronomy and Astrophysics, University of Chicago, Chicago IL 60637, USA}\\
}

\date{Accepted XXX. Received YYY; in original form ZZZ}

\pubyear{2022}

\begin{document}
\label{firstpage}
\pagerange{\pageref{firstpage}--\pageref{lastpage}}
\maketitle

\begin{abstract}
We present neutral atomic hydrogen (\ion{H}{i}) observations using the Robert C.\ Byrd Green Bank Telescope (GBT) along the lines of sight to 49 dwarf satellite galaxy candidates around eight Local Volume systems (M104, M51, NGC1023, NGC1156, NGC2903, NGC4258, NGC4565, NGC4631).\ We detect the \ion{H}{i} reservoirs of two candidates (dw0934+2204 and dw1238$-$1122) and confirm them as background sources relative to their nearest foreground host systems.\ The remaining 47 satellite candidates are not detected in \ion{H}{i}, and we place stringent $5\sigma$ upper limits on their \ion{H}{i} mass.\ We note that some (15/47) of our non-detections stem from satellites being occluded by their putative host's \ion{H}{i} emission.\ In addition to these new observations, we compile literature estimates on the \ion{H}{i} mass for an additional 17 satellites.\ We compare the \ion{H}{i} properties of these satellites to those within the Local Group, finding broad agreement between them.\ Crucially, these observations probe a ``transition'' region between $-10\gtrsim M_V \gtrsim -14$ where we see a mixture of gas-rich and gas-poor satellites and where quenching processes shift from longer timescales (i.e.\ via starvation) to shorter ones (i.e.\ via stripping).\ While there are many gas-poor satellites within this region, some are gas rich and suggests that the transition towards predominantly gas-rich satellites occurs at $L_V\sim10^{7}\lsun$, in line with simulations.\ The observations presented here are a key step toward characterizing the properties of dwarf satellite galaxies around Local Volume systems and future wide-field radio surveys with higher angular resolution (e.g.~WALLABY) will vastly improve upon the study of such systems.
\end{abstract}

\begin{keywords}
galaxies: dwarf -- galaxies: evolution -- (galaxies:) Local Group -- radio lines: galaxies
\end{keywords}

\section{Introduction}
Satellite galaxies provide a unique insight into the hierarchical galaxy formation and evolution process within the $\Lambda$CDM framework.\ Due to their proximity, the vast majority of detailed studies at low luminosities have been conducted with satellite dwarf galaxies in the Local Group.\ Several interesting trends have been discovered, some of which appear to be in tension with the current cosmological framework \citep[e.g.][]{2017bullockBoylanKolchin}, while others probe the environmental effects of the Milky Way and M31 on their satellites.\ Environmental trends in satellite dwarf galaxy properties are now well-established within the Local Group: low-mass ($M_*\lesssim10^6\msun$) dwarf satellites within the virial radius of the Milky Way or M31 are generally quenched and gas-poor, while those with higher-masses ($M_*\gtrsim10^8\msun$) or beyond the virial radius are generally star-forming and gas-rich \citep[][]{2009Grcevich, Spekkens2014, Putman2021}.\ Similarly, gas-rich and star-forming dwarf galaxies are ubiquitous in lower density environments \citep[i.e.\ the field][]{Huang2012,GehaNSAfielddwarfs}.\ Exceptions to these trends include quenched ``backsplash'' dwarf galaxies identified beyond the virial radius \citep[][]{Teyssier2012}, or ultra-faint dwarf galaxies plausibly quenched by reionization that appear in the field \citep[e.g.][]{Sand2022}.

Within the last decade, great strides have been made in constraining the quenching mechanisms of satellites galaxies from a theoretical perspective.\ The aforementioned environmental trends are also present in simulations of Milky Way-like ($M_{vir}\sim10^{12}\msun$) and Local Group-like systems \citep[][]{2015Fillingham,2016Fattahi,Wetzel2016,Simpson2018,2019garrisonkimmel,Akins2021,2021Karunakaran,2022Font}, regardless of their implementation (i.e.\ subgrid) of underlying astrophysical processes.\ Pushing these comparisons to lower masses with larger satellite samples is an important test for galaxy formation simulations, since lower-mass systems are more susceptible to these details of these processes due to their weaker gravitational potentials.\ 

Complimentary to these advances on the theoretical front, we are in an era of expanding studies of satellite dwarf galaxies beyond the Local Group that build upon the seminal works of \citet[][]{1993Zaritsky,1997Zaritsky}.\ These studies, whether via integrated light \citep[][]{2014MerritLSBsM101,2015Karachentsev,Bennet2017,JavanmardiDGSAT,CarlstenSBFM101,Muller2017,Smercina2018,GehaSAGA,ELVESI,2021Mao} or via resolved stars \citep[][]{Chiboucas2009,Chiboucas2013,Carlin16,PISCeS-CenAsats,Crnojevi__2019,Bennet2019,2020Bennet,2021MutluPakdil,2022MutluPakdil}, have discovered dozens of new satellites in nearby systems.\ These growing samples enable increasingly detailed comparisons to the Local Group satellite system.\ 

 \citet[][]{Carlsten2020} present a sample of 155 satellite candidates around 10 Local Volume ($<12$ Mpc) hosts detected in CFHT imaging.\ They then subsequently employed the surface brightness fluctuation (SBF) method to estimate distances to these candidates \citep[][]{Carlsten2021}, confirming 55 new satellites.\ While this Local Volume sample is near 100\% complete down to $M_V \gtrsim -9$ and $\mu_{0,V} \lesssim 26.5 \mathrm{mag\,arcsec^{-2}}$, its spatial coverage within the virial radius of the hosts is much lower compared to other surveys of Milky Way-like systems \citep[][]{GehaSAGA,2021Mao}.\ Nevertheless, the increased photometric completeness enables studies of the environmental effect on low-mass satellites by their hosts for the first time.\ 
 
A key complementary component to these wide-field optical satellite searches are observations of their neutral atomic hydrogen (\ion{H}{i}) content.\ Obtaining measurements of the satellite \ion{H}{i} content beyond the Local Group will place observational constraints on the environmental effects on these low-mass systems and also constrain the host-to-host scatter.\ As \ion{H}{i} is the initial fuel for star formation, its presence or lack thereof in satellites enables a better understanding of their past and future evolution.\ While the SBF distance method allows for a relatively robust estimate, there are occasions where it does not perform well \citep[e.g.\ for irregular morphologies,][]{2020Karunakarana,Carlsten2021}, and spectroscopic observations can help in these edge cases.\

Although, by-and-large, massive satellites are gas-rich and low-mass satellites are gas-poor within the Local Group, the threshold within this broad mass range at which the gas richness of the population transitions from low to high is only just beginning to be probed systematically \citep[i.e.][]{Carlsten2020, ELVESI}.\ This ``transition'' region lies above the stellar masses of the bulk of the Local Group satellites, but below the stellar masses of the bulk of the satellites of Milky Way-like systems that have insofar been detected in the Local Volume \citep[][]{GehaSAGA,2021Mao}.\ \ion{H}{i} observations of satellite candidates in this transition region therefore bridge the data gap between the Local Group and Local Volume while also constraining the mass dependence of the underlying quenching mechanisms at work.

In this paper we present new \ion{H}{i} observations of 49 dwarf satellite candidates around eight Local Volume hosts from the \citet[][]{Carlsten2020} sample with the Robert C.\ Byrd Green Bank Telescope\footnote{The Green Bank Observatory is a facility of the National Science Foundation operated under cooperative agreement by Associated Universities, Inc.} (GBT) and additionally compile 17 \ion{H}{i} measurements from the literature.\ With this study, we constrain the \ion{H}{i} gas content and gas richness of systems that reside in this aforementioned transition region for the first time.\ In addition, we lay the foundation for more comprehensive studies of the \ion{H}{i} properties of satellites around massive hosts in the Local Volume and beyond, while also highlighting some potentially interesting trends that will be solidified with future expanded studies.\ 

The structure of this paper is as follows.\ In Section \ref{sec:sample} and \ref{sec:obs}, we describe our sample selection and our \ion{H}{i} observations.\ We present our derived and compiled \ion{H}{i} results in Section \ref{sec:results}, along with a brief discussion of the properties of their optical counterparts and a comparison to the Local Group satellites.\ In Section \ref{dissandsum}, we briefly discuss this work in a broader context and provide our summary.\

\section{Sample Selection}\label{sec:sample}
We select our \ion{H}{i} follow-up sample from the Local Volume survey conducted by \citet[][hereafter \citetalias{Carlsten2020}]{Carlsten2020} and \citet[][hereafter \citetalias{Carlsten2021}]{Carlsten2021}.\ A total of 155 satellite candidates around 10 Local Volume hosts were presented in \citetalias{Carlsten2020} with subsequent SBF distance estimates presented in \citetalias{Carlsten2021}.\ The distance estimates were used to classify satellite candidates as ``confirmed'', ``possible'' (unconstrained), or ``background'' with respect to their putative hosts.\ A total of 55 of the \citetalias{Carlsten2020} candidates were confirmed as satellites, 48 classified as possible, and the remaining 49 classified as background systems.\ Based on the mock dwarf injection/recovery testing presented in \citetalias{Carlsten2021}, the sample is considered to be near 100\% complete for $M_V \gtrsim -9$, $\mu_{0,V} \lesssim 26.5 \mathrm{mag\,arcsec^{-2}}, \mathrm{and \, r_{eff} >4}''$.\ However, we note that the spatial coverage of the \citetalias{Carlsten2020} sample is not as complete.\ Only 6 of the 9 hosts studied here have greater than 70\% coverage within a 150 kpc projected radius.\ We keep this caveat in mind for our interpretation.

For our \ion{H}{i} follow-up sample, we select all satellites brighter than $M_V = -9.5$ ($M_*\sim10^6\msun$) that are classified as ``confirmed'' or ``possible''.\ We opted for this selection limit primarily to minimize the amount of observing time that would be required, however, it also ensures that we are well within the photometric completeness limit of the sample.\ This selection criterion produces a sample of 66 satellite candidates (48 confirmed, 18 possible), 17 of which have \ion{H}{i} measurements (either detections or upper limits) in the literature.\ We list basic properties of the studied sample in Table \ref{tab:maintable}.\ Throughout this work, we assume that the distances to the satellites are the same as their hosts unless otherwise stated.\

\section{Observations and Data Reduction}\label{sec:obs}
We performed a total of $\sim72$ hours of observations (projects GBT20A-576 and GBT21A-388, PI:Karunakaran) with the GBT using the L-band receiver and the VErsatile GBT Astronomical Spectrometer (VEGAS) along the lines-of-sight to 49 satellite candidates from \citetalias{Carlsten2020} and \citetalias{Carlsten2021} without literature \ion{H}{i} detections (see Table~\ref{tab:maintable}).\ GBT20A-576 focused on the brighter ($M_V \leq -11$ mag) subset of confirmed or possible satellite candidates, while GBT21-388 focused on the fainter ($-9.5 \geq M_V > -11$) targets.\ Our observing strategy for each subset differed.\ For the brighter subset, we used VEGAS in Mode 10 which provides a relatively narrow bandwidth (23.44MHz, $\sim5000\kms$).\ Given the robustness of the SBF technique at higher luminosity, a wider bandpass would not have benefited the search for the \ion{H}{i} reservoirs of these systems and would likely have been more detrimental in terms of radio frequency interference (RFI).\ Conversely, for the fainter subset, we used VEGAS in Mode 7 which provides a wider bandpass (100MHz, $\sim21000\kms$).\ This wider bandwidth affords the ability to search for potential \ion{H}{i} signals along the LOS out to velocities of $14000\kms\,(\sim200\mathrm{Mpc})$.\ While we could have centered our bandpass to probe a greater velocity range, we have found, from previous observations, that there is strong, intermittent RFI at higher (lower) velocities (frequencies) that can severely affect these deep observations.\ To estimate the required sensitivities (i.e.\ RMS noise levels) to detect the \ion{H}{i} reservoirs of these sources, we used their V-band luminosities with an assumed gas-richness of $\mhi/\LV = 1 \msun/\lsun$ across $25\kms$ channels.\ This gas-richness limit generally separates gas-poor satellites and gas-rich field dwarf galaxies within the Local Group \citep[][]{Spekkens2014} and is also $\sim2\sigma$ below the scaling relations of \citet[][]{Bradford2015}.\

We follow the standard procedure to calibrate the raw GBT spectra using \textsc{getps} in GBTIDL\footnote{https://gbtidl.nrao.edu/index.shtml} as presented in \citet{2020Karunakarana,2020Karunakaranb}.\ As part of this procedure, we flag and replace narrowband RFI with local noise values in a given 5s integration (a data dump) and remove entire data dumps that are affected by broadband RFI, specifically the 1.38GHz GPS-L3 signal (see \citealt{2020Karunakarana} for more details).\ Following these standard RFI excision measures, we found that several of the calibrated spectra were affected by unforeseen and infrequent RFI that resulted in broadband artifacts.\ Therefore, we opted to remove these affected data dumps ($\sim15-30\%$ of them depending on the target) and repeat the calibration process.\ For these reasons, we were unable to reach the desired $\mhi/\LV$ for several of our targets.\ 

\begin{figure*}
	\includegraphics[width=\textwidth]{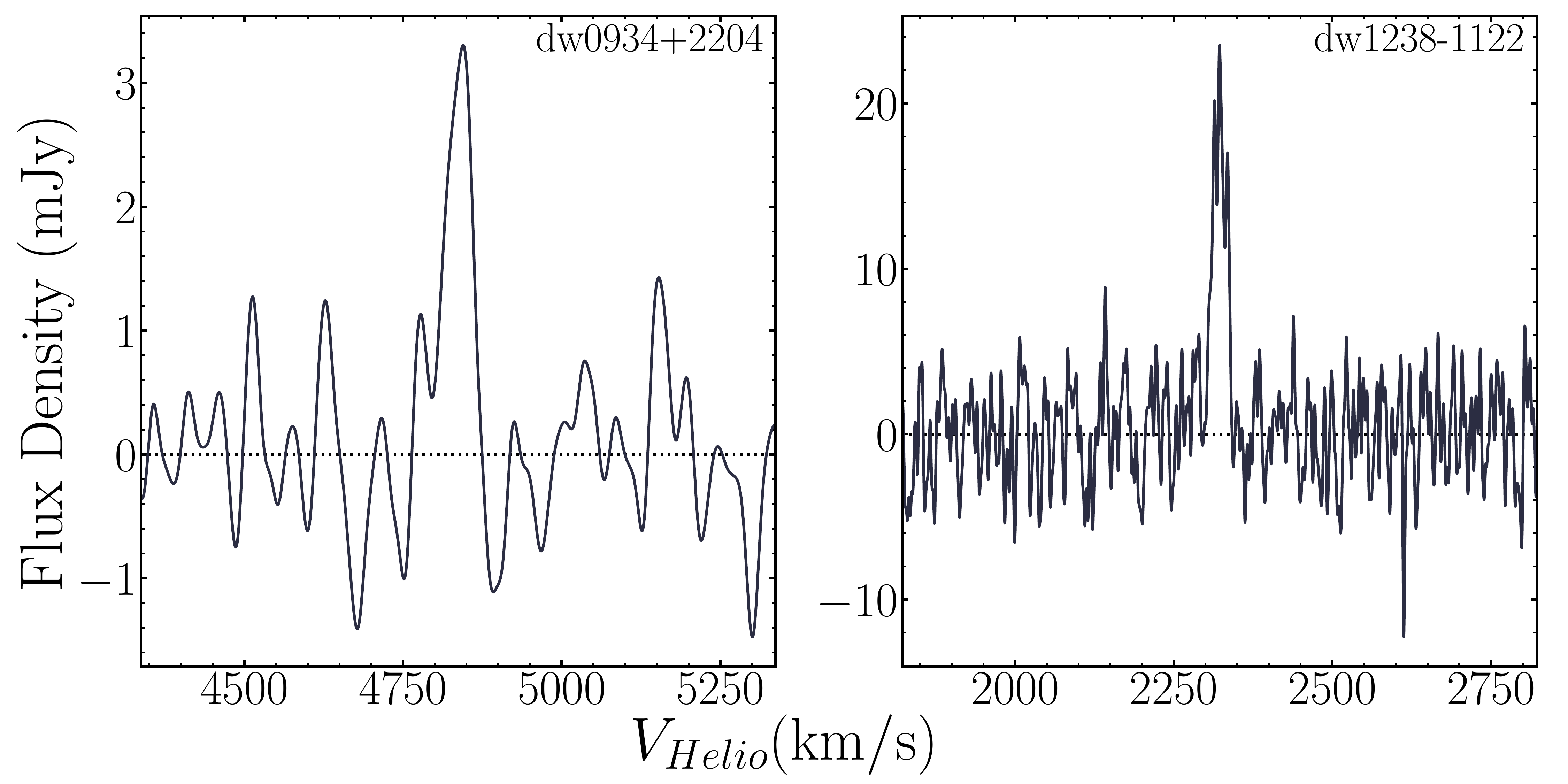}
    \caption{\ion{H}{i} spectra for dw0934+2204 and dw1238$-$1122 from our GBT observations at the resolution, $\Delta$V, listed alongside their derived \ion{H}{i} properties in Table \ref{table:detectiontable}.\ The horizontal dotted line shows a flux density of 0 mJy.\ These detections confirm that the satellite candidates are in fact in the background of their putative hosts, and are therefore field dwarfs.\ }
    \label{fig:detections}
\end{figure*}

We visually search for potentially significant \ion{H}{i} emission in the calibrated, RFI-excised spectra that we smooth to various velocity resolutions $5\,\kms < \Delta V < 50\,\kms$.\ In Table \ref{tab:maintable}, we list representative RMS noise values ($\sigma_{50}$) for all of our targets in the emission-free regions of each spectrum at a velocity resolution $\Delta V=50\kms$.\ We detect \ion{H}{i} along the LOS to two of our targets, dw0934+2204 and dw1238$-$1122.\ We show their \ion{H}{i} spectra in Figure \ref{fig:detections} and list their derived properties in Table \ref{table:detectiontable}.\ For the remaining 47 targets, we estimate upper limits on $\mhi$ and $\mhi/\LV$, and list them in Table \ref{tab:maintable}.\ We note that $\sim$30\% (15/47) of the non-detections have \ion{H}{i} emission from their host's or a nearby neighbour's \ion{H}{i} disk.\ While this leads to less stringent constraints on whether or not they are truly gas-rich satellites, we treat these systems as non-detections and discuss this issue in more detail in Section \ref{subsec:nondet} and in the context of their optical properties in Section \ref{subsec:optcolmorph}.

\section{Results}\label{sec:results}
\subsection{\ion{H}{i} Detections}
Prior to deriving their properties, we first confirm that we have correctly associated our two \ion{H}{i} detections with their targeted optical counterparts and not nearby interlopers.\ Given the well-characterized response pattern of the GBT beam (FWHM$\sim9'$) at 1.420GHz down to $\approx-30$dB \citep{GBTbeam}, we can search for potential interlopers and confirm the association of these detections to the satellite candidates.\ We performed a search through NED\footnote{The NASA/IPAC Extragalactic Database (NED) is operated by the Jet Propulsion Laboratory, California Institute of Technology, under contract with the National Aeronautics and Space Administration.} within a radius of 30$'$ and within $\pm500\kms$ of the systemic velocity of the \ion{H}{i} detection.\ We also visually searched through the Legacy Survey Viewer\footnote{https://www.legacysurvey.org/viewer} and Pan-STARRS cutouts\footnote{http://ps1images.stsci.edu/cgi-bin/ps1cutouts} for potential gas-rich sources (i.e.\ relatively blue, late-type or irregular galaxies) within 30$'$.\ We find no such sources in our search, strongly suggesting that the \ion{H}{i} detections are the counterparts to the two satellite candidates in our sample.\

We follow the methods described in \citet{2020Karunakarana} to derive the properties of our two \ion{H}{i} detections.\ We first estimate the systemic velocities, $V_{sys}$, and velocity widths, $\wfty{}$, by performing a linear fit at each edge of the \ion{H}{i} profile between 15\% and 85\% of the peak \ion{H}{i} flux.\ From these fits, we find the velocity that corresponds to 50\% of the peak flux at each edge and their average provides $V_{sys}$, while their difference provides $\wfty$.\ We correct $\wfty$ for instrumental broadening and cosmological redshift, resulting in a corrected velocity width $\wftyc$.\ The adopted 50\% uncertainty on the instrumental broadening correction (see \citealt{Springob2005}) dominates the uncertainties of both $V_{sys}$ and $\wftyc$.\ These values and their uncertainties are listed in Table \ref{table:detectiontable}.\

Before we estimate $\mhi$ for our detections, we first must estimate their distances.\ We use our derived $V_{sys}$ values together with the Hubble-Lema\^{i}tre law assuming $\mathrm{H_0} = 70\,\kms\,\mathrm{Mpc^{-1}}$ to estimate their distances and we assume distance uncertainties of 5 Mpc.\ Both of these sources are in the distant background of their putative host galaxies and within the Hubble flow with distances of 69 Mpc for dw0934+2204 and 33 Mpc for dw1238$-$1122.\ Additionally, as we described above, we find no massive companions near these dwarfs.\ This is generally consistent with their ``possible'' association classification in \citetalias{Carlsten2021}, as well as the note made by those authors regarding the challenge of deriving S\'{e}rsic models for and estimating SBF distances for dw1238$-$1122.\

With distance estimates $D_{HI}$ in-hand, we now compute the \ion{H}{i} mass $\mhi$ using the standard relation assuming an optically thin gas \citep{Haynes1984}
\begin{equation}M_{HI}=2.356\times 10^{5}(D_{HI})^{2}S_{HI} \, \mathrm{\msun},\end{equation} 
where $D_{HI}$ is in Mpc and $S_{HI}$ is the \ion{H}{i} flux in Jy $\kms$ computed by integrating over the \ion{H}{i} profile.\ We estimate the uncertainty on the \ion{H}{i} mass following the methods of \cite{Springob2005} and including the 5 Mpc distance uncertainty in quadrature.\ We list these derived properties and their uncertainties in Table \ref{table:detectiontable}.\ As part of our aforementioned search for interlopers, we found no massive systems that could be possible hosts for these two background systems and consider them to be dwarf galaxies in the field.

\subsubsection{\textit{GALEX} UV Photometry of new \ion{H}{i} detections}
We perform aperture photometry of deep\footnote{i.e.\ exposure times > 1000s, with the exception of a single AIS depth ~100s FUV tile} archival \textit{GALEX} UV imaging for the two detections in our sample to complement their \ion{H}{i} derived properties.\ We follow the curve-of-growth method described in \citet{2021Karunakaran} to find the optimal radius at which fluxes are measured.\ To estimate the background and noise, we place 1000 equal-sized background apertures in $15'\times15'$ cutout images centered on the dwarf and take the mean as the background value and the standard deviation as the noise.\ We compute AB apparent magnitudes using the standard equations \citep[][]{2007MorrisseyGALEX} (see Table~\ref{table:uvtable}) and correct for foreground extinction using $E(B-V)$ from \citet{2011Schlaflydust} with $R_{NUV} = 8.2, R_{FUV} = 8.24$ \citep[][]{2007wyder}.\ Using these extinction-corrected magnitudes and $D_{HI}$ with the relations from \citet{2006IglesiasParamo}, we estimate star formation rates $SFR_{NUV}$ and $SFR_{FUV}$.\ Together with these SFRs, we estimate approximate gas-consumption timescales for these field dwarf galaxies and find that, in addition to their \ion{H}{i} properties, they are similar to the broader field dwarf galaxy population \citep[e.g.][]{Huang2012}.\ We list all of these derived properties along with their \textit{GALEX} tile names in Table \ref{table:uvtable}.

\subsubsection{Optical Properties of new \ion{H}{i} Detections}\label{sec:optprophi}
We briefly discuss the optical properties of our two new \ion{H}{i} detections.\ dw0934+2204 is an LSB dwarf galaxy in the field with a relatively smooth morphology, as indicated by the 'dE' classification from \citetalias{Carlsten2020}, and is blue in colour, $g-r\sim0.3$, akin to many other LSB dwarf galaxies in low-density environments \citep[e.g.][]{2021Tanoglidis}.\ On the other hand, dw1238$-$1122, has optical properties near the threshold criteria ($\mu_{0,g}\gtrsim 24 \mathrm{mag\,arcsec^{-2}}$ and $\mathrm{r_{eff} \gtrsim1.5\mathrm{\,kpc}}$) for an Ultra-Diffuse Galaxy (UDG, \citealt[][]{2015vandokkum}) with $\mu_{0,g}\sim 23.7 \mathrm{mag\,arcsec^{-2}}$ and $\mathrm{r_{eff}\sim2.3}$ kpc\footnote{\citetalias{Carlsten2020} fit S\'{e}rsic profiles to derive effective surface brightnesses and here we have estimated $\mu_{0,g}$ assuming n=1.\ Of course, $\mu_{0,g}$ will vary depending on the true n for this system.}.\ Furthermore, its relatively narrow velocity width, $\wftyc=14\pm4\kms$, is also consistent with the broader UDG population \citep[e.g.][]{Leisman2017, 2020Karunakaranb,2022Poulain}.\

\subsection{\ion{H}{i} non-detections}\label{subsec:nondet}
For the remaining 47 sources in our sample observed with the GBT, we find no obvious \ion{H}{i} counterparts.\ We place $5\sigma$ upper limits on their \ion{H}{i} masses assuming their host distances and $\sigma_{50}$ from Table \ref{tab:maintable} together with a modified version of Eq.(1): 
\begin{equation} \mhilim=5.89\times 10^{7}D_{host}^{2}({\sigma_{50}})\, \mathrm{\msun}.\end{equation}
We list $\mhilim$ and $\mhi/\LV$ upper limits in column 13 of Table \ref{tab:maintable}.\ It is reasonable to assume that these satellites are at the distances of their hosts given the particular strength of the SBF distance estimation method for relatively red, early-type systems such as our non-detections (see Section \ref{subsec:optcolmorph}) and, by contrast, exceptions to this trend for relatively blue, irregular systems, such as our \ion{H}{i} detections.\

\begin{figure*}
	\includegraphics[width=\textwidth]{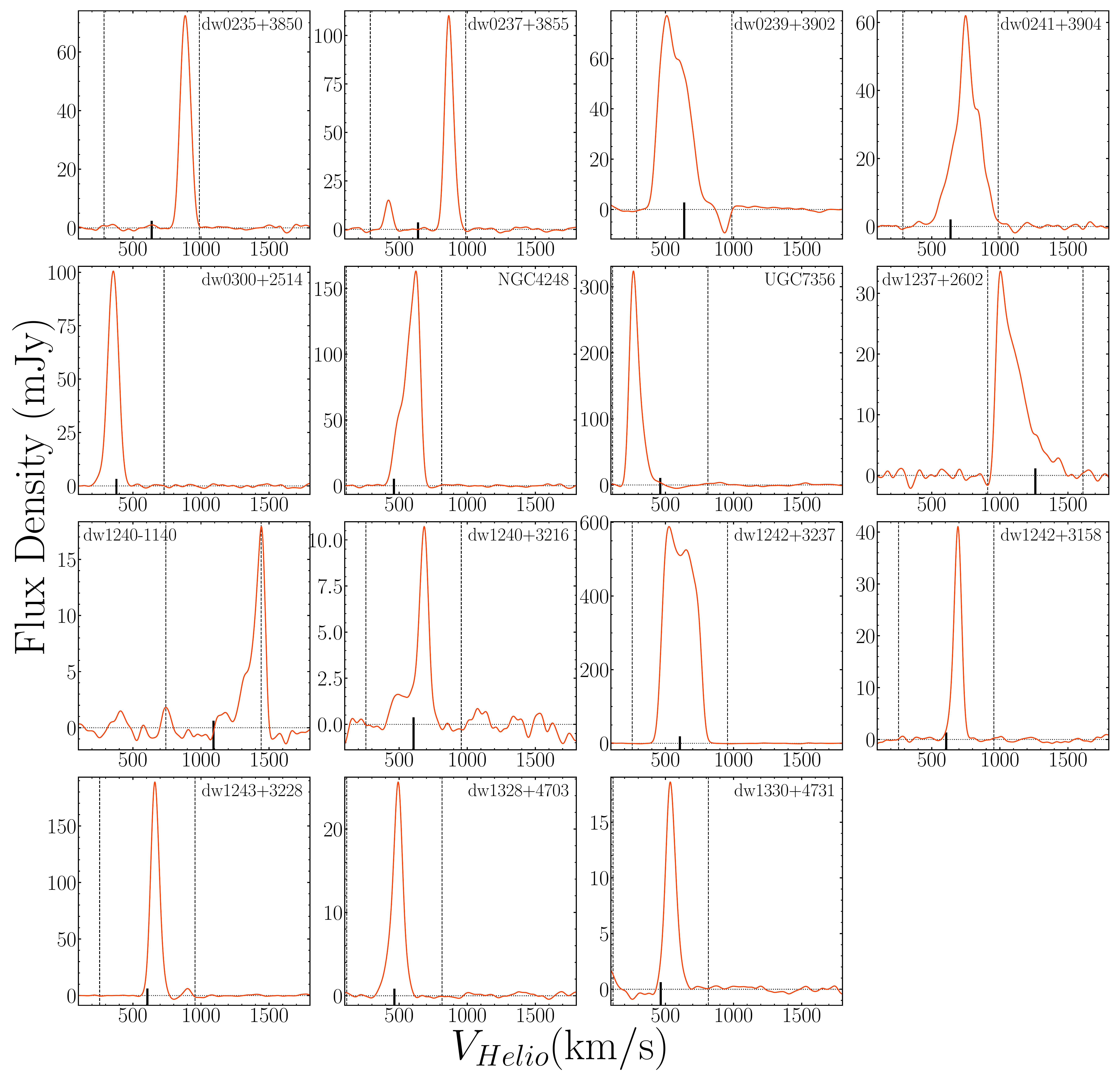}
    \caption{\ion{H}{i} spectra for our non-detections that contain \ion{H}{i} emission contamination from their hosts.\ The vertical dashed lines indicate the velocity range in which we expect the host \ion{H}{i} emission to dominate and the short solid line shows the host systemic velocity.\ The horizontal dotted line shows a flux density of 0 mJy.\ We also indicate these affected satellites in Table \ref{table:detectiontable} with a * next to their names.\ }
    \label{fig:obscuredspectra}
\end{figure*}

Some of these sources are either confused by their host's or a nearby, more massive satellite's \ion{H}{i} emission and we mark their names in Table \ref{tab:maintable} with a * symbol.\ In Figure \ref{fig:obscuredspectra} we show the spectra of the 15 obscured targets in our observed sample.\ The vertical dashed lines show the approximate velocity range we expected their host \ion{H}{i} emission to cover, i.e.\ their systemic velocity (short, solid vertical lines) $\pm350\kms$.\ From this figure, we can see that these systems have strong \ion{H}{i} contamination from their hosts.\ We can also see that in several of these cases the entire velocity range is not contaminated as our observations have likely only partially detected the contaminating disk due to the GBT beam response pattern.\ That is to say, the strength and shape of the contaminating \ion{H}{i} emission depends on the host \ion{H}{i} disk's orientation and distance from the GBT pointing center.\ These cases allow us to further constrain the velocity space that the satellite could reside in within the host's gravitational reach.\ So, while it is possible that a few of these sources may indeed have \ion{H}{i} reservoirs of their own, we were unable to discern them based on the available data and higher spatial resolution \ion{H}{i} data may provide more insight in this regard.\ We return to this issue in Section \ref{sec:opthi}.\
\subsection{Literature \ion{H}{i} Measurements}
In addition to the new GBT observations of 49 satellites, we compile \ion{H}{i} observations for 17 satellites from the literature.\ We include whether or not the source has a detected \ion{H}{i} counterpart, its integrated flux estimate, corresponding source papers, and $\mhi$ or upper limit in Table \ref{tab:maintable}.\ We estimate $\mhi$ and $\mhi/\LV$ assuming their host distances.\ 15 of these sources have confirmed \ion{H}{i} reservoirs.\ We derive an upper-limit for NGC4627 because the detection listed by \citet[][]{Wolfinger2013} is a case of confusion with its host's (NGC4631's) \ion{H}{i} emission.\ In contrast, our derived upper limit for UGC5086 stems from VLA observations with higher spatial resolution than the original detection, distinguishing the \ion{H}{i} disc of NGC2903 from the lack of emission at the position of UGC5086 \citep[][]{Irwin2009}.\ Unsurprisingly, all of the sources from the literature are bright with $M_V \lesssim -11$ relative to the broader sample.\ This suggests that dedicated \ion{H}{i} observations of fainter systems are required to push beyond what is presently available in the literature.\

\subsection{Comparisons of optical and \ion{H}{i} properties}\label{sec:opthi}
Here, we make brief comparisons of the newly derived and compiled \ion{H}{i} properties of the satellite candidates with their optical properties to gain more insight into the interplay between various tracers.\ We also make general comparisons between the Local Volume sample studied here and the satellites from the Local Group.

\subsubsection{Optical Colours and Morphologies}\label{subsec:optcolmorph}
\begin{figure}
	\includegraphics[width=\columnwidth]{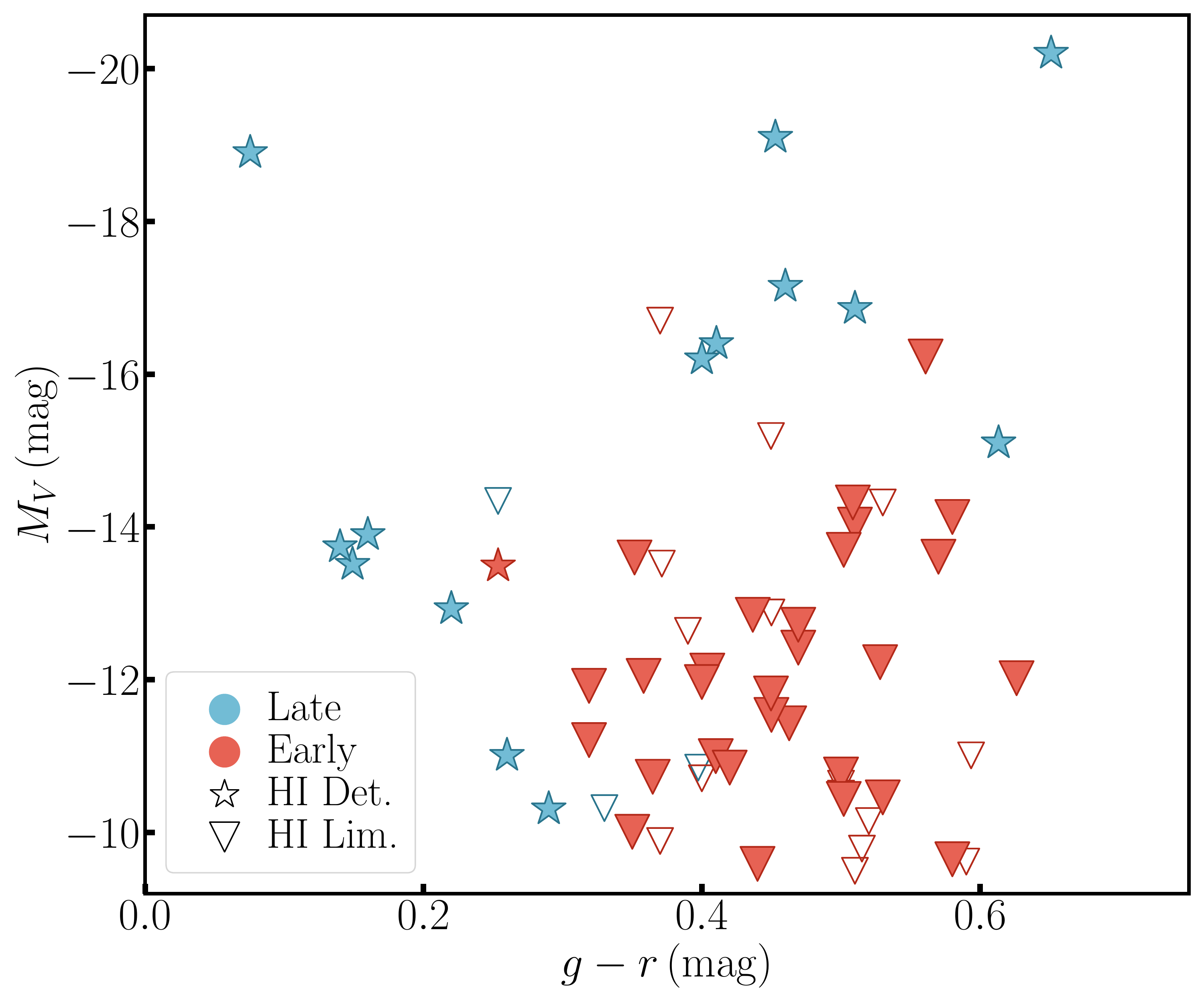}
    \caption{Comparison of $M_V$ versus $g-r$ colour for the Local Volume sample.\ The symbol shapes represent \ion{H}{i} detections (stars) and non-detections (inverted triangles), while the symbol colours correspond to the general morphological classification provided by \citetalias{Carlsten2020}, blue for late-types and red for early-types.\ We show satellites with \ion{H}{i} detections or stringent non-detections (i.e.\ $\mhi/\LV \leq2$) as filled symbols, whereas satellites with weaker limits on $\mhi/\LV$ or those which were obscured by their hosts' \ion{H}{i} emission are represented by open symbols.\ }
    \label{fig:MV_gr}
\end{figure}

 We first investigate the relationship between a satellite's optical colour, morphological class, and whether or not it has been detected in \ion{H}{i}.\ In Figure \ref{fig:MV_gr} we show $M_V$ as a function of $g-r$ for satellites with \ion{H}{i} detections or satellites with relatively stringent non-detections (i.e.\ $\mhi/\LV \leq2;\,\sim1\sigma$ off the relations of \citealt[][]{Bradford2015}).\ These systems are represented by filled symbols, whereas satellites with weaker limits on $\mhi/\LV$ or were obscured by their hosts' \ion{H}{i} emission are represented by open symbols.\ We separate the satellites into broad ``Late'' (blue) and ``Early'' (red) classes based on the morphological classifications in \citetalias{Carlsten2020}.\ Satellites that are detected in \ion{H}{i} are shown as stars, while non-detections are shown as inverted triangles.\ We have used the $g-r$ values from \citetalias{Carlsten2021} and we convert any $g-i$ colours listed in that work to $g-r$ using Equation (1) in \citet[][]{ELVESII}.\ We note that there are four satellites (IC239, dw0240+3903, NGC4656, and NGC5195) that do not have a listed $g-r$ colour in \citetalias{Carlsten2021}.\ For three of these sources, we convert their $B-V$ colours listed in HyperLeda \citep[][]{HyperLEDA} to $g-r$ using the relations provided by \citet{Jestervband}.\ For NGC4656, we estimate $g-r$ using the SDSS photometry from \citet{2012schectman}.\ Finally, we convert these SDSS $g-r$ colours to CFHT $g-r$ using the relation derived by \citetalias{Carlsten2020} (see their Equation 2).\ 

We focus our comparison on the satellites with \ion{H}{i} detections and stringent non-detections, revealing an interesting and potentially insightful trend.\ As we move toward fainter satellites (i.e.\ $M_V \gtrsim-14$), they fall towards redder colours, are not detected in \ion{H}{i}, and are predominantly early-type in their morphology.\ The one exception to this is dw0240+3854 ($g-r \sim0.25$, $M_V\sim-13.5$) which is detected in \ion{H}{i}, has a relatively blue optical colour, and through visual inspection is clearly visible in \textit{GALEX} NUV and FUV imaging\footnote{See \href{https://www.legacysurvey.org/viewer/cutout.jpg?ra=40.1372&dec=38.9010&layer=galex&pixscale=0.15&size=400}{Legacy Survey Viewer} for a colour composite} despite its early-type morphology.\ While there are cases of host \ion{H}{i} confusion or RFI-related issues leading to weak limits on $\mhi/\LV$, we can see that the aforementioned trend is broadly true for these other systems and supports the gas-poor nature of the majority of them.\ We discuss this trend further in the following section.\

\subsubsection{Comparisons to the Local Group Satellites}
\begin{figure}

	\includegraphics[width=\columnwidth]{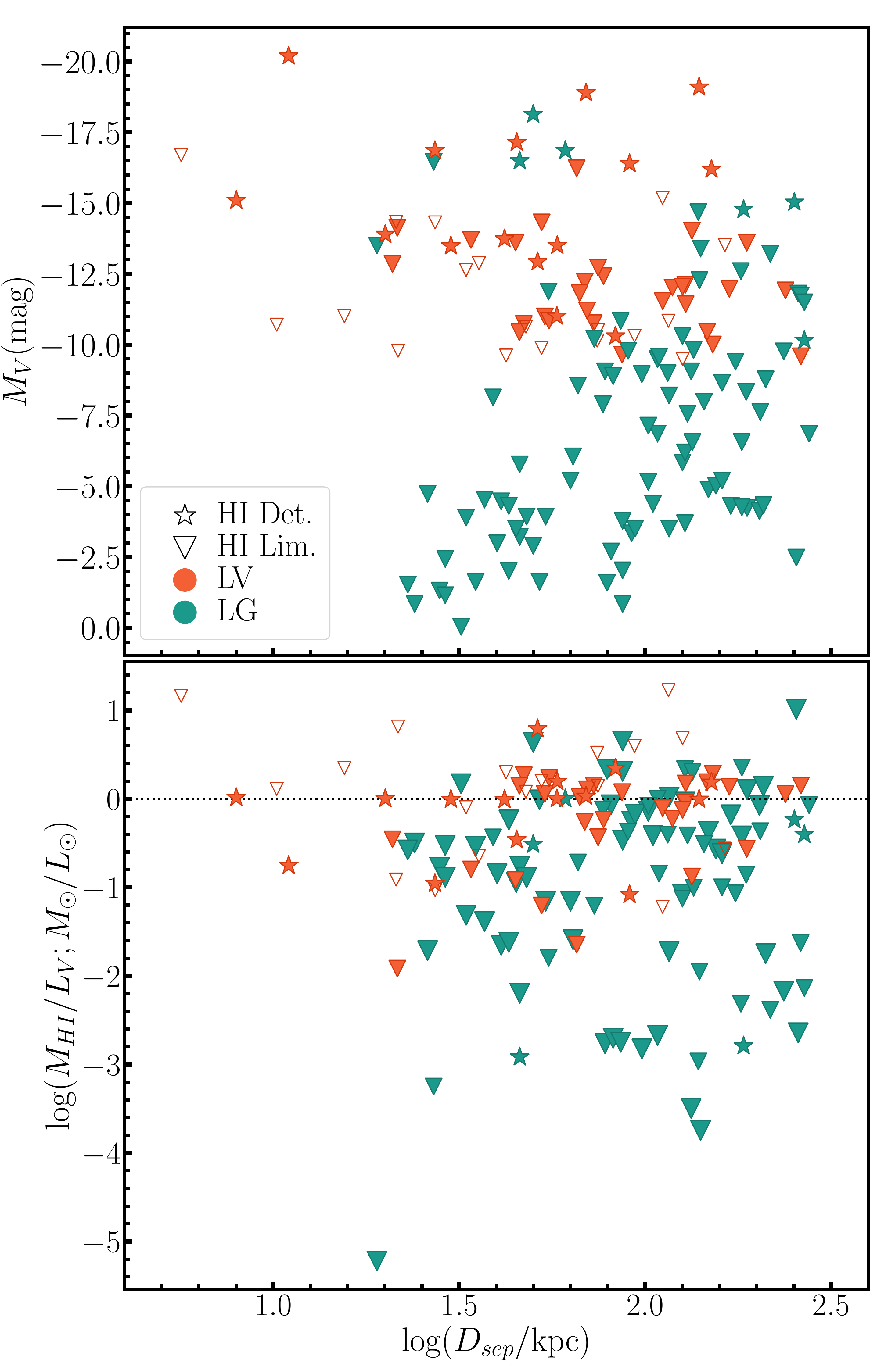}
    \caption{Comparisons of the V-band absolute magnitudes (top) and \ion{H}{i} mass to V-band luminosity ratios, $\mhi/\LV$, (bottom) as a function of separation for the Local Volume (orange symbols) and Local Group (green symbols) samples.\ Stars represent satellites with \ion{H}{i} detections, while inverted triangles show those with \ion{H}{i} mass upper limits.\ Filled symbols are \ion{H}{i} detections or stringent non-detections, while open symbols are satellites with host-obscured spectra or with weak upper-limits on $\mhi/\LV$.\ The Local Volume separations show their projected distances, whereas the Local Group separations are their true distances from either the Milky Way or M31 listed in \citet[][]{Putman2021}.\ The horizontal dotted line in the bottom panel indicates $\mhi/\LV$ = 1 $\msun/\lsun$ which generally separates gas-rich field dwarfs and gas-poor satellites.\ }
    \label{fig:comparisons}
\end{figure}

We now turn to the Local Group and make comparisons with the sample in this work.\ In Figure \ref{fig:comparisons} we show Local Group (green, \citealt[][]{Putman2021}) and Local Volume (orange) satellite $M_V$ and log$(\mhi/\LV)$ as a function of separation from their hosts.\ We note that we exclude the NGC1156 system from this figure as it is not in the same luminosity/mass regime as the Milky Way and M31 \citepalias{Carlsten2021}.\ We show satellites with \ion{H}{i} detections as stars and \ion{H}{i} upper limits as inverted triangles.\ As in Figure \ref{fig:MV_gr}, we show \ion{H}{i} detections and stringent non-detections as filled symbols, while open symbols represent satellites with host-obscured spectra or weak upper-limits on $\mhi/\LV$.\ From the top panel of Figure \ref{fig:comparisons} we can see that our \ion{H}{i} observations are beginning to probe further down the satellite luminosity function into the region of gas-poor Local Group dwarfs.\ Similarly, we are beginning to probe a similar parameter space as the Local Group satellites in terms of gas-richness ($\mhi/\LV$) as seen in the bottom panel of Figure \ref{fig:comparisons}.\ However, the proximity of the bulk of the Local Group satellites leads to much more stringent limits overall.\ For reference we include a horizontal dotted line indicating $\mhi/\LV$ = 1 $\msun/\lsun$ which separates gas-poor satellites and gas-rich field dwarf galaxies.\ Considering both panels in Figure \ref{fig:comparisons} together suggests that we are now starting to probe a transition region between $(-10\gtrsim M_V \gtrsim -14)$ where we see a mixture of gas-rich and gas-poor satellites.\ We discuss this in more detail with respect to results from simulations in the following section.\

\section{Discussion and Summary}\label{dissandsum}
We have presented new \ion{H}{i} observations of 49 satellites around eight Local Volume hosts using the GBT.\ We detect \ion{H}{i} in two systems (dw0934+2204 and dw1238$-$1122) that confirm they are in the background of the Local Volume hosts near which they project.\ These two systems have \ion{H}{i} and star-forming properties consistent with the field dwarf galaxy population \citep[e.g.][]{Huang2012} and one of which has properties near the threshold of UDGs (see Section \ref{sec:optprophi}).

For the remaining 47 sources in our sample we set $5\sigma$ upper-limits on their \ion{H}{i} mass.\ In addition to these new observations, we compile \ion{H}{i} measurements from the literature for 17 satellites.\ We compare the \ion{H}{i} properties of these 64 satellites around Local Volume hosts to the satellites in the Local Group (see Figure \ref{fig:comparisons}).\ We find that the gas richnesses, $\mhi/\LV$, for the Local Volume satellites are broadly similar to those of the Local Group.\ Furthermore, with this sample of satellites that push to even fainter optical luminosities, we are beginning to probe a transition region between $-10\gtrsim M_V \gtrsim -14$.\ Dwarf satellites above this threshold are predominantly star-forming and gas rich, while those below it are quenched and gas poor.\ This trend is more clearly seen in Figure \ref{fig:MV_gr} where we show only the Local Volume sample and distinguish satellites by their optical morphology and whether or not they were detected in \ion{H}{i}.\ This is an interesting and insightful consistency that we also see in the Local Group (Figure \ref{fig:comparisons}; \citealt[][]{Putman2021}) and is the first observational demonstration of such a trend around other Milky Way-like systems.\ While many of the satellites in this transition region are gas-poor, some are gas-rich.\ This result suggests that the transition between predominantly gas-poor and gas-rich satellites occurs at $L_V \sim10^{7}\lsun$, in line with predictions from simulations \citep[][]{2015Fillingham}.\ Furthermore, this consistency suggests that similar quenching processes typically invoked for dwarf galaxies in the Local Group are likely to be at play in these other systems.\ Similarly, more massive satellites have been shown to be quenched and/or gas-poor in accordingly higher density environments such as groups and clusters \citep[][]{Brown2015,Brown2017,Jones2020}, reaffirming the greater susceptibility of lower mass halos to environmental effects leading to their eventual quenching as seen in hydrodynamical simulations \citep[][]{Fillingham2016, 2019garrisonkimmel,2022Samuel}.\ Compiling existing and obtaining new \ion{H}{i} observations would allow for quantitative comparisons to theoretical predictions beyond the qualitative initial comparisons discussed here.

\begin{figure*}

	\includegraphics[width=\textwidth]{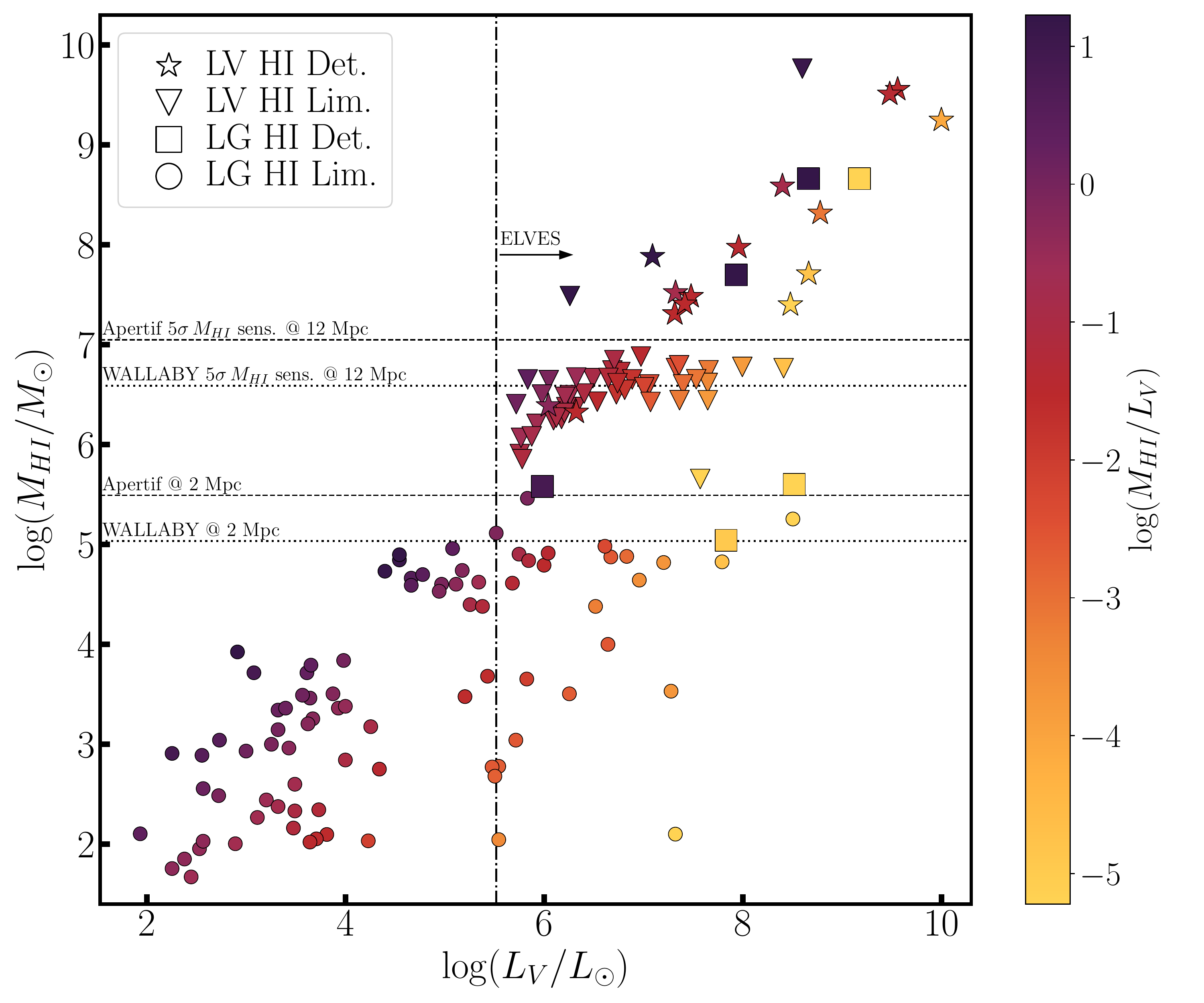}
    \caption{\ion{H}{i} mass as a function of V-band luminosity for the Local Volume sample from this work (stars and inverted triangles) and the dwarf galaxies in the Local Group within 300 kpc of the Milky Way or M31 (squares and circles).\ The colours of the symbols show the logarithm of $\mhi/\LV$.\ The horizontal dashed lines shows the $5\sigma$ \ion{H}{i} lower limits that probed by the Apertif survey at distances of 2 (lower line) and 12 (upper line) Mpc.\ Similarly, the horizontal dotted line shows the $5\sigma$ \ion{H}{i} lower limits probed by the WALLABY survey.\ It should be noted that these wide-field surveys are at much higher spatial resolutions, $\sim15\arcsec$ and $\sim30\arcsec$ beam widths for Apertif and WALLABY, respectively.\ The vertical dash-dotted line shows the completeness limit of the ELVES survey.}
    \label{fig:mhi_lv}
\end{figure*}

While the observations presented in this work are an important step towards understanding the \ion{H}{i} properties of other satellites systems in the Local Volume, we briefly consider the parameter space that will be probed by upcoming \ion{H}{i} surveys.\ In Figure \ref{fig:mhi_lv}, we show log$(\mhi)$ as a function of log$(\LV)$ for the Local Volume (stars and triangles) and Local Group (squares and circles) satellites coloured by their gas richness, log$(\mhi/\LV)$.\ The horizontal dashed lines and dotted lines show the estimated minimum $\mhi$ that will be probed by the upcoming Apertif survey data releases \citep[][Hess et al.\ in prep.]{Apertif} and upcoming WALLABY \citep[][]{WALLABY} survey, respectively, at distances of 2 Mpc (lower lines) and 12 Mpc (upper lines).\ Furthermore, we note that these estimates assume unresolved $5\sigma$ sources with velocity widths of $50\kms$.\ The aforementioned transition region can be seen ($\LV\sim10^{6-7.5}\lsun$) with a mix of \ion{H}{i} detections (stars and squares) and non-detections (triangles and circles).\ While we are able to reach similar satellite gas richness limits with the deep observations presented in this work to those in the Local Group, confirming this transition region requires a larger sample of satellites and \ion{H}{i} observations.\

More quantitative comparisons may be made using the Exploration of Local VolumE Satellites (ELVES) Survey \citep[][]{ELVESI}.\ The ELVES sample extends the one used in this work and consists of over 300 confirmed satellites around 30 Local Volume hosts with more uniform spatial coverage within 300 kpc and similar photometric completeness, vertical dashed-dotted line in Figure \ref{fig:mhi_lv}.\ This sample will populate the aforementioned transition region and with additional \ion{H}{i} constraints, we can place statistically significant constraints on this region.\ Furthermore, the additional spatial coverage will enable studies of gas-richness as a function of radial separation.\ The Apertif and WALLABY survey areas includes 8 and 18 of the ELVES systems, respectively.\ Of these 24 systems with Apertif and WALLABY coverage, 8 were studied in this work albeit with significantly less spatial completeness.\ So, while we were able to identify some potentially interesting trends, such as the one between colour, morphology, and \ion{H}{i} emission from Figure \ref{fig:MV_gr}, the increased sample size will solidify their validity.\ These \ion{H}{i} surveys will not only provide great sensitivity but their spatial resolution (Apertif$\sim15''$, WALLABY $\sim30''$) will reduce the occurrence of host \ion{H}{i} confusion, may resolve the \ion{H}{i} distributions in the most massive satellites, and possibly detect the remnants of past interactions (i.e.\ \ion{H}{i} streams).\

There is still much to be done until these upcoming surveys are fully on-line and/or their data analysed.\ With this in mind, we have initiated additional follow-up surveys to characterize the \ion{H}{i} and star-forming properties of satellites galaxies in the Local Universe.\ This initial follow-up effort aims to set additional groundwork for what future wide-field \ion{H}{i} surveys, like WALLABY, will tell us.\

\begin{landscape}
    \begin{table}
        \centering
        \caption{Properties of Confirmed and Possible satellites from the Local Volume sample.\ Col.\ (1): Dwarf satellite names.\ Sources indicated with a $^{\dagger}$ are detections from this work and those with a * have contaminating \ion{H}{i} emission from their host or neighbour.\ Cols.\ (2) and (3): J2000 Right Ascension and Declination.\ Cols.\ (4) and (5): Putative host galaxy and host distance.\ Col.\ (6): V-band absolute magnitude taken from \citet{Carlsten2021}.\ (1).\ Col.\ (7): Association, either confirmed or possible based on \citet{Carlsten2021}.\ Col.\ (8): \ion{H}{i} source abbreviations are as follows: V77 = \citet{vanAlbada1977}, S84 = \citet{Sancisi1984}, B03 = \citet{Braun2003}, D05 = \citet{Dahlem2005}, I09 = \citet{Irwin2009}, W13 = \citet{Wolfinger2013}, C15 = \citet{Courtois2015}, H18 = \citet{Haynes2018}, K20 =\citet[][]{2020Karunakaranb}, K22 = this work.\ Col.\ (9): \ion{H}{i} detection or non-detection.\ Cols.\ (10) and (11): Integrated flux and RMS noise at a velocity resolution of 50 $\kms$.\ Col.\ (12): Observing time with the GBT.\ Cols.\ (13) and (14): Logarithm of \ion{H}{i} mass and \ion{H}{i} mass to V-band luminosity ratio.\ In the case of non-detections, $5\sigma$ upper limits are reported.\ }
        \begin{tabular}{cccccccccccccc}
             \hline
             Name &	$\alpha$ & $\delta$ & Host & $D_{Host}$ & $\mathrm{M_V}$ & C21 & \ion{H}{i} & \ion{H}{i} & Int.\ Flux & $\sigma_{50}$ & Time &$\mathrm{log}(\mhi)$ & $\mhi/\LV$ \\
             & (H:M:S) & (D:M:S) & & (Mpc) & (mag) & Assoc.\ & Source & Det? & (Jy $\kms$) & (mJy) & (hours) & $\mathrm{1og}(\msun)$  & $(\msun/\lsun)$ \\
             (1) & (2) & (3) & (4) & (5) & (6) & (7) & (8) & (9) & (10) & (11) & (12) & (13) & (14)\\
             \hline
dw0233+3852 & 02:33:42.70 & 38:52:20.10 & NGC1023 & 10.4 & $-11.92$ & C & K22 &  & -- & 0.87 & 0.3 & <6.74 & <1.13 \\
dw0235+3850* & 02:35:54.20 & 38:50:10.30 & NGC1023 & 10.4 & $-13.52$ & C & K22 &  & -- & 0.92 & 0.3 & <6.77 & <0.27 \\
IC239 & 02:36:28.10 & 38:58:08.50 & NGC1023 & 10.4 & $-19.1$ & C & B03 & Y & 140.9 & -- & -- & 9.56 & 0.99 \\
dw0237+3855* & 02:37:18.60 & 38:55:59.20 & NGC1023 & 10.4 & $-15.19$ & C & K22 &  & -- & 0.94 & 0.3 & <6.78 & <0.06 \\
dw0237+3836 & 02:37:39.40 & 38:36:01.20 & NGC1023 & 10.4 & $-12.12$ & C & K22 &  & -- & 0.84 & 0.3 & <6.73 & <0.90 \\
dw0238+3805 & 02:38:41.00 & 38:05:06.50 & NGC1023 & 10.4 & $-13.6$ & P & K22 &  & -- & 0.98 & 0.3 & <6.79 & <0.27 \\
dw0239+3926 & 02:39:19.90 & 39:26:02.10 & NGC1023 & 10.4 & $-12.42$ & C & K22 &  & -- & 0.7 & 0.3 & <6.65 & <0.58 \\
dw0239+3902* & 02:39:47.00 & 39:02:50.40 & NGC1023 & 10.4 & $-9.79$ & C & K22 &  & -- & 0.7 & 1.5 & <6.65 & <6.54 \\
UGC2157 & 02:40:25.00 & 38:33:46.90 & NGC1023 & 10.4 & $-16.4$ & C & C15 & Y & 1.44 & -- & -- & 7.40 & 0.1 \\
dw0240+3854 & 02:40:33.00 & 38:54:01.40 & NGC1023 & 10.4 & $-13.49$ & C & S84 & Y & 0.8 & -- & -- & 7.31 & 0.98 \\
dw0240+3903 & 02:40:37.10 & 39:03:33.60 & NGC1023 & 10.4 & $-15.1$ & C & S84 & Y & 3.7 & -- & -- & 7.97 & 1.03 \\
dw0240+3922 & 02:40:39.60 & 39:22:45.10 & NGC1023 & 10.4 & $-13.51$ & C & S84 & Y & 1.3 & -- & -- & 7.52 & 1.57 \\
dw0241+3904* & 02:41:00.40 & 39:04:20.60 & NGC1023 & 10.4 & $-14.34$ & C & K22 &  & -- & 0.87 & 0.3 & <6.74 & <0.12 \\
UGC2165 & 02:41:15.50 & 38:44:38.90 & NGC1023 & 10.4 & $-16.23$ & C & K22 &  & -- & 0.91 & 0.3 & <6.76 & <0.02 \\
dw0241+3829 & 02:41:54.20 & 38:29:53.60 & NGC1023 & 10.4 & $-10.85$ & P & K22 &  & -- & 4.81 & 1.9 & <7.49 & <16.85 \\
dw0243+3915 & 02:43:55.00 & 39:15:20.70 & NGC1023 & 10.4 & $-11.43$ & P & K22 &  & -- & 0.73 & 0.5 & <6.66 & <1.49 \\
dw0300+2514* & 03:00:17.80 & 25:14:56.00 & NGC1156 & 7.6 & $-10.66$ & P & K22 &  & -- & 1.05 & 0.5 & <6.55 & <2.34 \\
dw0301+2446 & 03:01:32.20 & 24:46:59.40 & NGC1156 & 7.6 & $-10.76$ & P & K22 &  & -- & 2.31 & 0.3 & <6.89 & <4.68 \\
dw0930+2143 & 09:30:40.00 & 21:43:27.10 & NGC2903 & 8 & $-11.01$ & C & I09 & Y & 0.14 & -- & -- & <6.32 & 1.00 \\
UGC5086 & 09:32:48.80 & 21:27:56.20 & NGC2903 & 8 & $-14.13$ & C & I09 &  & -- & -- & -- & <5.66 & <0.01 \\
dw0934+2204$^{\dagger}$ & 09:34:22.00 & 22:04:53.90 & NGC2903 & 8 & $-14.96$ & P & K22 & Y & 0.18 & 0.34 & 1.6 & 8.29 & 2.48 \\
NGC4248* & 12:17:50.20 & 47:24:33.40 & NGC4258 & 7.2 & $-16.86$ & C & V77 & Y & 4.19 & -- & -- & 7.71 & 0.11 \\
LVJ1218+4655 & 12:18:11.20 & 46:55:02.00 & NGC4258 & 7.2 & $-12.93$ & C & W13 & Y & 6.26 & -- & -- & 7.88 & 6.19 \\
dw1219+4743 & 12:19:06.20 & 47:43:49.30 & NGC4258 & 7.2 & $-11.00$ & C & K22 &  & -- & 0.78 & 0.2 & <6.38 & <1.14 \\
UGC7356* & 12:19:09.00 & 47:05:23.90 & NGC4258 & 7.2 & $-14.32$ & C & K22 &  & -- & 1.36 & 0.2 & <6.62 & <0.09 \\
dw1220+4922 & 12:20:14.40 & 49:22:51.60 & NGC4258 & 7.2 & $-9.59$ & P & K22 &  & -- & 0.26 & 2.9 & <5.91 & <1.41 \\
dw1220+4649 & 12:20:54.90 & 46:49:48.40 & NGC4258 & 7.2 & $-10.76$ & C & K22 &  & -- & 0.78 & 0.5 & <6.37 & <1.41 \\
dw1223+4739 & 12:23:46.20 & 47:39:32.70 & NGC4258 & 7.2 & $-11.54$ & C & K22 &  & -- & 0.88 & 0.2 & <6.43 & <0.78 \\
dw1233+2535 & 12:33:11.00 & 25:35:55.20 & NGC4565 & 11.9 & $-11.97$ & P & K22 &  & -- & 0.84 & 0.2 & <6.84 & <1.37 \\
dw1233+2543 & 12:33:18.40 & 25:43:35.10 & NGC4565 & 11.9 & $-10.01$ & P & K22 &  & -- & 0.19 & 3.8 & <6.21 & <1.92 \\
dw1234+2531 & 12:34:24.20 & 25:31:20.20 & NGC4565 & 11.9 & $-14.03$ & C & K22 &  & -- & 0.54 & 0.5 & <6.65 & <0.13 \\
dw1234+2618 & 12:34:57.60 & 26:18:50.80 & NGC4565 & 11.9 & $-10.32$ & P & K22 &  & -- & 0.53 & 3.5 & <6.65 & <3.96 \\
dw1235+2616 & 12:35:22.30 & 26:16:14.20 & NGC4565 & 11.9 & $-10.15$ & P & K22 &  & -- & 0.38 & 3.7 & <6.50 & <3.32 \\
NGC4562 & 12:35:34.70 & 25:51:01.30 & NGC4565 & 11.9 & $-17.15$ & C & H18 & Y & 6.22 & -- & -- & 8.32 & 0.34 \\
IC3571 & 12:36:20.00 & 26:05:03.50 & NGC4565 & 11.9 & $-13.90$ & C & D05 & Y & 0.91 & -- & -- & 7.48 & 1.01 \\
dw1236+2634 & 12:36:58.60 & 26:34:42.80 & NGC4565 & 11.9 & $-9.50$ & P & K22 &  & -- & 0.3 & 4.0 & <6.40 & <4.84 \\
dw1237+2602* & 12:37:01.20 & 26:02:09.60 & NGC4565 & 11.9 & $-12.64$ & C & K22 &  & -- & 0.91 & 0.2 & <6.88 & <0.80 \\
dw1237+2605 & 12:37:26.80 & 26:05:08.70 & NGC4565 & 11.9 & $-10.85$ & P & K22 &  & -- & 0.37 & 1.8 & <6.49 & <1.71 \\
dw1237+2637 & 12:37:42.80 & 26:37:27.60 & NGC4565 & 11.9 & $-10.46$ & P & K22 &  & -- & 0.23 & 3.6 & <6.29 & <1.54 \\
 
             \hline
        \end{tabular}
        \label{tab:maintable}
    \end{table}
\end{landscape}

\begin{landscape}
    \begin{table}
        \centering
        \contcaption{}
        \begin{tabular}{cccccccccccccc}
            \hline
             Name &	$\alpha$ & $\delta$ & Host & $D_{Host}$ & $\mathrm{M_V}$ & C21 & \ion{H}{i} & \ion{H}{i} & Int.\ Flux & $\sigma_{50}$ & Time &$\mathrm{log}(\mhi)$ & $\mhi/\LV$ \\
             & (H:M:S) & (D:M:S) & & (Mpc) & (mag) & Assoc.\ & Source & Det? & (Jy $\kms$) & (mJy) & (hours) & $\mathrm{1og}(\msun)$ & $(\msun/\lsun)$ \\
             (1) & (2) & (3) & (4) & (5) & (6) & (7) & (8) & (9) & (10) & (11) & (12) & (13) & (14)\\
             \hline
dw1239+3230 & 12:39:05.00 & 32:30:16.50 & NGC4631 & 7.4 & $-10.31$ & C & K20 & Y & 0.19 & -- & -- & 6.39 & 2.22 \\
dw1239+3251 & 12:39:19.60 & 32:51:39.30 & NGC4631 & 7.4 & $-9.65$ & C & K22 &  & -- & 0.22 & 2.6 & <5.85 & <1.19 \\
dw1240+3216* & 12:40:53.00 & 32:16:55.90 & NGC4631 & 7.4 & $-10.64$ & C & K22 &  & -- & 0.56 & 0.6 & <6.26 & <1.20 \\
dw1240+3247 & 12:40:58.50 & 32:47:25.00 & NGC4631 & 7.4 & $-13.61$ & C & K22 &  & -- & 0.86 & 0.2 & <6.44 & <0.12 \\
dw1241+3251 & 12:41:47.10 & 32:51:27.30 & NGC4631 & 7.4 & $-13.74$ & C & H18 & Y & 1.98 & -- & -- & 7.41 & 0.98 \\
NGC4627 & 12:41:59.70 & 32:34:26.20 & NGC4631 & 7.4 & $-16.7$ & C & W13 &  & -- & -- & -- & <9.76 & <14.57 \\
dw1242+3237* & 12:42:06.20 & 32:37:18.70 & NGC4631 & 7.4 & $-10.71$ & C & K22 &  & -- & 0.64 & 0.5 & <6.32 & <1.30 \\
dw1242+3158* & 12:42:31.40 & 31:58:09.20 & NGC4631 & 7.4 & $-10.51$ & C & K22 &  & -- & 0.58 & 0.8 & <6.27 & <1.40 \\
dw1243+3228* & 12:43:24.80 & 32:28:55.30 & NGC4631 & 7.4 & $-12.88$ & C & K22 &  & -- & 0.83 & 0.2 & <6.43 & <0.23 \\
NGC4656 & 12:43:57.70 & 32:10:05.30 & NGC4631 & 7.4 & $-18.9$ & C & H18 & Y & 250.18 & -- & -- & 9.51 & 1.07 \\
dw1237$-$1125 & 12:37:11.60 & $-$11:25:59.30 & M104 & 9.55 & $-12.02$ & C & K22 &  & -- & 0.59 & 0.4 & <6.50 & <0.60 \\
dw1238$-$1122$^{\dagger}$ & 12:38:33.70 & $-$11:22:05.10 & M104 & 9.55 & $-12.6$ & P & K22 & Y & 0.57 & 0.71 & 0.4 & 8.17 & 1.36 \\
dw1239$-$1159 & 12:39:09.10 & $-$11:59:12.20 & M104 & 9.55 & $-11.21$ & C & K22 &  & -- & 0.6 & 0.8 & <6.51 & <1.28 \\
dw1239$-$1143 & 12:39:15.30 & $-$11:43:08.10 & M104 & 9.55 & $-13.70$ & C & K22 &  & -- & 0.74 & 0.4 & <6.60 & <0.16 \\
dw1239$-$1113 & 12:39:32.70 & $-$11:13:36.00 & M104 & 9.55 & $-12.23$ & C & K22 &  & -- & 0.66 & 0.4 & <6.55 & <0.54 \\
dw1239$-$1120 & 12:39:51.50 & $-$11:20:28.70 & M104 & 9.55 & $-10.73$ & C & K22 &  & -- & 0.56 & 0.6 & <6.48 & <1.84 \\
dw1239$-$1144 & 12:39:54.90 & $-$11:44:45.50 & M104 & 9.55 & $-12.85$ & C & K22 &  & -- & 0.74 & 0.4 & <6.60 & <0.35 \\
dw1240$-$1118 & 12:40:09.40 & $-$11:18:49.80 & M104 & 9.55 & $-14.32$ & C & K22 &  & -- & 0.52 & 0.4 & <6.44 & <0.06 \\
dw1240$-$1140* & 12:40:17.60 & $-$11:40:45.70 & M104 & 9.55 & $-11.01$ & C & K22 &  & -- & 0.87 & 0.5 & <6.67 & <2.22 \\
dw1241$-$1131 & 12:41:02.80 & $-$11:31:43.70 & M104 & 9.55 & $-10.44$ & C & K22 &  & -- & 0.33 & 1.9 & <6.24 & <1.41 \\
dw1241$-$1153 & 12:41:12.10 & $-$11:53:29.70 & M104 & 9.55 & $-11.82$ & C & K22 &  & -- & 0.87 & 0.3 & <6.67 & <1.04 \\
dw1241$-$1155 & 12:41:18.70 & $-$11:55:30.80 & M104 & 9.55 & $-12.72$ & C & K22 &  & -- & 0.7 & 0.4 & <6.57 & <0.37 \\
dw1242$-$1116 & 12:42:43.80 & $-$11:16:26.00 & M104 & 9.55 & $-12.05$ & P & K22 &  & -- & 0.77 & 0.4 & <6.61 & <0.75 \\
dw1328+4703* & 13:28:24.70 & 47:03:54.80 & M51 & 8.6 & $-9.62$ & P & K22 &  & -- & 0.27 & 2.6 & <6.07 & <1.99 \\
NGC5195 & 13:29:59.60 & 47:15:58.10 & M51 & 8.6 & $-20.2$ & C & C15 & Y & 101.56 & -- & -- & 9.25 & 0.18 \\
dw1330+4731* & 13:30:33.90 & 47:31:33.10 & M51 & 8.6 & $-9.89$ & P & K22 &  & -- & 0.28 & 2.6 & <6.08 & <1.61 \\
NGC5229 & 13:34:03.00 & 47:54:49.80 & M51 & 8.6 & $-16.2$ & C & C15 & Y & 22.23 & -- & -- & 8.59 & 1.54 \\
            \hline
        \end{tabular}
    \end{table}
\end{landscape}
 
\begin{table*}
    \begin{center}
        
        \caption{\ion{H}{i} Properties of Dwarf Satellite Candidates with \ion{H}{i} detections.\ Cols.\ (2) and (3): Velocity resolution and RMS noise at $\Delta_V$.\ Col.\ (4): Systemic velocity of the \ion{H}{i} detection.\ Col.\ (5): Velocity width corrected for instrumental and redshift broadening.\ Col.\ (6): Integrated \ion{H}{i} flux density.\ Col.\ (7): Distance calculated from $V_{sys}$ in col.\ (4) using the Hubble-Lema\^{i}tre law assuming $\mathrm{H_0} = 70\,\kms\,\mathrm{Mpc^{-1}}$.\ Cols.\ (8) and (9): Logarithm of the V-band luminosity and \ion{H}{i} mass.\ Col.\ (10): \ion{H}{i} mass to V-band luminosity ratio.\ }
        \label{table:detectiontable}
        \begin{tabular}{cccccccccc}
        \hline
        {Name} & {$\Delta V$} & {$\sigma_{\Delta V}$} & {$V_{sys}$} & {$W_{50,c}$} & {$S_{HI}$} & {$D_{HI}$} & {log($\LV$)} &
        {log($\mhi$)} & {$({\mhi}/{\LV})$}  \\
        {} & {($\kms$)} & {(mJy)} & {($\kms$)} & {($\kms$)} & {(Jy$\,\kms$)} & {(Mpc)} &
        {(log[$\lsun$])} & 
        {(log[$\msun$])} & {$({\msun}/{\lsun})$} \\
        {(1)} & {(2)} & {(3)} & {(4)} & {(5)} &
        {(6)} & {(7)} & {(8)} &
        {(9)} & {(10)}\\
        \hline
        dw0934+2204 & 20 & 0.6 & 4837  $\pm$ 2 & 39 $\pm$ 3 & 0.18 $\pm$ 0.05 & 69$\pm$5 & 7.90 $\pm$ 0.08 & 8.29 $\pm$ 0.13 & 2.48 $\pm$ 0.84 \\
        dw1238$-$1122 & 15 & 1.3 & 2322 $\pm$ 3 & 14 $\pm$ 4 & 0.57 $\pm$ 0.08 & 33$\pm$5 & 8.04 $\pm$ 0.26 & 8.17 $\pm$ 0.11 & 1.36 $\pm$ 0.89 \\
        \hline
        \end{tabular}
    \end{center}
\end{table*}

\begin{table*}
    \begin{center}
        
        \caption{\textit{GALEX} UV Properties of Dwarf Satellite Candidates with \ion{H}{i} detections.\ Cols.\ (2) and (3): Apparent NUV and FUV magnitudes corrected for foreground extinction.\ Cols.\ (4) and (5): Logarithm of star formation rates (SFR) calculated from NUV and FUV luminosities using the relations of \citet[][]{2006IglesiasParamo}.\ Col.\ (6): Approximate gas consumption timescale in Gyr calculated using the FUV SFR and $\mhi$ listed in col.\ (9) of Table \ref{table:detectiontable}.\ Cols.\ (7) and (8): The NUV and FUV tile names.\ }
        \label{table:uvtable}
        \begin{tabular}{cccccccccc}
        \hline
        {Name} & {$m_{NUV}$} & {$m_{FUV}$} & {$\mathrm{log}(SFR_{NUV})$} & {$\mathrm{log}(SFR_{FUV})$} & {$\mathrm{T_{cons}}$} & {NUV Tile} & {FUV Tile}\\
        {} & {(mag)} & {(mag)} & {($\mathrm{log}(M_{\odot}\,\mathrm{yr^{-1}})$)} & {($\mathrm{log}(M_{\odot}\,\mathrm{yr^{-1}})$)} & ({$\mathrm{Gyr}$}) & {} & {}\\
        {(1)} & {(2)} & {(3)} & {(4)} & {(5)} & {(6)} & {(7)} & {(8)}\\
        \hline
        dw0934+2204 & 20.9 $\pm$ 0.4 & 21.5 $\pm$ 0.5 & $-2.11$ $\pm$ 0.18 & $-2.54$ $\pm$0.20 & $\sim80$ & MISGCSN3\_23812\_0193 & AIS\_192\_1\_39\\
        dw1238$-$1122 & 20.1 $\pm$ 0.4 & 20.4 $\pm$ 0.4 & $-2.41$ $\pm$ 0.21 & $-2.74$ $\pm$ 0.21 & $\sim60$ & NGA\_NGC4594 & NGA\_NGC4594\\
        \hline
        \end{tabular}
    \end{center}
\end{table*}

\section*{Acknowledgements}
We thank Kelley M.\ Hess for useful discussions regarding the Apertif survey.\ AK acknowledges financial support from the State Agency for Research of the Spanish Ministry of Science, Innovation and Universities through the "Center of Excellence Severo Ochoa" awarded to the Instituto de Astrof\'{i}sica de Andaluc\'{i}a (SEV-2017-0709) and through the grant POSTDOC$\_$21$\_$00845 financed from the budgetary program 54a Scientific Research and Innovation of the Economic Transformation, Industry, Knowledge and Universities Council of the Regional Government of Andalusia.\ KS acknowledges support from the Natural Sciences and Engineering Research Council of Canada (NSERC).\ BMP is supported by an NSF Astronomy and Astrophysics Postdoctoral Fellowship under award AST2001663.\ Research by DC is supported by NSF grant AST-1814208.\  DJS acknowledges support from NSF grants AST-1821967 and 1813708.

\section*{Data Availability}
The raw and reduced \ion{H}{i} spectra from the GBT used in this work as well as the derived properties listed in Tables \ref{tab:maintable}, \ref{table:detectiontable}, \ref{table:uvtable} may be shared upon request to A.\ Karunakaran.\

\bibliographystyle{mnras}
\bibliography{references}

\bsp	
\label{lastpage}
\end{document}